\documentclass[aps,prd,amsmath,groupedaddress,twocolumn,showp
acs]{revtex4}
\usepackage{amssymb}

\newcounter{gplfmfgraph}

\newcommand{\smalltext}{\footnotesize}
\newcommand{\be}{\begin{equation}}
\newcommand{\ee}{\end{equation}}
\newcommand{\bea}{\begin{eqnarray}}
\newcommand{\eea}{\end{eqnarray}}
\newcommand{\bc}{\begin{center}}
\newcommand{\ec}{\end{center}}
\newcommand{\bi}{\begin{itemize}}
\newcommand{\ei}{\end{itemize}}
\newcommand{\bis}{\begin{itemize} \smalltext}
\newcommand{\eis}{\end{itemize}}
\newcommand{\biss}{\begin{itemize} \smalltext}
\newcommand{\eiss}{\end{itemize}}
\newcommand{\bfr}{\begin{frame}}
\newcommand{\efr}{\end{frame}}
\newcommand{\order}{{\cal O}}

\newcommand{\e}{{\rm e}}

\newcommand{\psib}{\overline{\psi}}

\newcommand{\B}{{\cal B}}
\newcommand{\Bz}{{\cal B}_\zeta}

\newlength{\boxitlen}

\newlength{\shrinkboxlen}

\newcommand{\eq}[1]{Eq.~(\ref{#1})}
\newcommand{\nl}{\nonumber \\}
\newcommand{\bearray}{\begin{eqnarray}}
\newcommand{\eearray}{\end{eqnarray}}

\newcommand{\omittext}[1]{}

\newlength{\minuslength}
\newlength{\digitlength}
\newcommand{\cS}{\mathcal{S}}

\begin{document}
\title{On the Absence of $\order(a)$ Errors in Staggered-Quark 
Discretizations}

\author{G. Peter Lepage}
\affiliation{Laboratory for Elementary-Particle Physics, Cornell 
University, Ithaca, NY 14853, USA}

\date{23 March 2007 (updated Nov. 2011)}

\begin{abstract}
We demonstrate that the $\order(a)$~taste mixing exhibited in standard 
textbook presentations of staggered quarks is an artifact of the 
particular definition of the flavor fields in those presentations, and 
has nothing to do with the underlying precision of staggered-quark 
actions, despite continuing comments to the contrary in the current 
literature. To illustrate this point we introduce a new 
coordinate-space definition of the flavor fields that suppresses the 
$\order(a)$ term by two additional powers of~$a$. In fact there are no 
errors at all from this mechanism. The only source of taste mixing 
comes from the exchange of highly-virtual gluons and enters in 
$\order(a^2)$. We review the idiosyncrasies of Symanzik improvement for 
naive/staggerd-quark actions, and show how these results follow from 
that program.
\end{abstract}


\maketitle

Improved versions of the staggered-quark action are proving highly 
effective for precise numerical simulations of full 
QCD~\cite{ratio-paper,alphas-paper,mb-paper,fd,DtoK}. These formalisms 
have the unusual property that a single quark field creates several 
different flavors or ``tastes'' of quark, each with the same mass. This 
multiplication of quark degrees of freedom is unphysical but can be 
remedied to the extent that the different tastes decouple from each 
other~\cite{hisq-paper,sharpe-paper,Kronfeld:2010aw,Donald:2011if}. 
Unfortunately, despite almost thirty years of study, there remains 
confusion about the origins and size of taste mixing in these theories. 
It is critically important to resolve this issue given the central role 
played by taste mixing in the error analysis of staggered-quark results 
and given the theoretical issues raised by the presence of taste 
mixing~\cite{sharpe-paper,Kronfeld:2010aw,Donald:2011if}.

Much of this confusion is caused by what have become standard textbook 
treatments of the flavor structure of staggered 
quarks~\cite{textbooks}. These treatments, which ignore interactions, 
re-express the free staggered-quark action in terms of a set of 
interpolating fields, with one field for each taste of quark. There are 
two standard choices for this set of fields: the coordinate-space 
flavor basis, and the momentum-space flavor basis. When rewritten in 
terms of the canonical coordinate-space flavor fields, the 
staggered-quark action has an~$\order(a)$ taste-mixing term, of the 
schematic form
\be \label{stagg-a-term}
 a\sum_\mu\psib\gamma_5\otimes\xi_5\xi_\mu \partial^2_\mu \psi,
\ee
where our notation is from~\cite{hisq-paper}.
The momentum-space basis, however, has no taste-mixing at all in the 
noninteracting case. Notwithstanding this discrepancy, several authors 
continue to assume that taste-mixing interactions occur in~$\order(a)$.

In fact, as we discuss in this note, the $\order(a)$~term in the 
coordinate-space analysis is highly misleading. It arises only because 
the coordinate-space flavor fields break one of the symmetries of the 
underlying staggered-quark theory: translation invariance on the 
original lattice. The $\order(a)$~term is needed to cancel errors 
introduced by the flavor fields. It has nothing to do with the 
underlying theory. Generally it is a bad idea to break the underlying 
symmetries when forming interpolating fields, and there is no need to 
do so here. The momentum-space flavor basis, for example, does not 
violate translation symmetry. Indeed the coordinate-space basis is 
never used in practical calculations; all such calculations are much 
more easily framed in terms of the momentum-space 
basis~\cite{app-hisq-paper}.

To underscore the artificial origins of the canonical $\order(a)$~term 
we will construct here a new coordinate-space flavor basis for which 
the taste-mixing term is $\order(a^3)$ rather than~$\order(a)$. Since 
the underlying theory is the same in both cases, neither the 
$\order(a)$~term nor the $\order(a^3)$~term can be relevant to anything 
other than representations of the free theory in terms of the 
corresponding, flawed operators\,---\,in other words, neither is 
relevant to anything important. The momentum-space analysis gets the 
right answer: taste-mixing is impossible in the noninteracting case. As 
we will show, it is forbidden by the symmetries of the underlying 
staggered-quark theories.

Contrary to what is sometimes asserted, there is only one source for 
taste mixing and that is the exchange of highly virtual gluons between 
quarks~\cite{lepage1-paper,sinclair-paper,lepage2-paper,hisq-paper}. 
Obviously this mechanism cannot be analyzed in the noninteracting 
theory, and therefore any analysis of the noninteracting 
staggered-quark theory has nothing to say about real taste mixing. Here 
we will review what is known about real taste mixing. In particular we 
will review the arguments, some now quite old, for why these effects 
and all other finite-$a$ errors enter first only in~$\order(a^2)$. We 
will also discuss how these errors show up in the spectrum of the 
staggered-quark Dirac operator, which is important for understanding 
vacuum polarization and for understanding instanton effects.

To address the $\order(a)$~term, we only need examine free quarks since 
that term appears in the canonical coordinate-space analysis of the 
free theory. Rather than use the staggered-quark discretization 
directly, we will work with the formally equivalent but intuitively 
much simpler ``naive'' discretization of the quark 
action~\cite{hisq-paper}:
\be
	\sum_x
	\psib(x)\gamma_\mu\frac{\psi(x+a\hat\mu)-\psi(x-a\hat\mu)}{2a}
	+ \psib(x)m\psi(x)
\ee
The gamma matrices and other conventions used here are described 
in~\cite{hisq-paper}.

This action is invariant under a ``doubling transformation''
\be \label{dsym}
\psi(x) \,\to\, \B_\zeta(x)\,\psi(x) \quad\quad
\psib(x) \,\to\, \psib(x)\,\Bz^\dagger(x)
\ee
where
\bea \label{Bz-def}
\Bz(x) &\equiv& \gamma_{\,\overline{\zeta}}\,(-1)^{\zeta\cdot x/a} \nl
&\propto&
\prod_\rho\,\left(\gamma_5\gamma_\rho\right)^{\zeta_\rho}\,\,
\exp(i\,x\cdot\zeta\,\pi/a),\label{B}
\eea
and $\zeta$ is a four-vector of 0s~and~1s (i.e., 
$\zeta_\mu\in\mathbb{Z}_2$)~\cite{hisq-paper}. The ``conjugate'' 
$\overline\zeta$ to~$\zeta$ is (see Appendix~A in~\cite{hisq-paper}):
\begin{equation}
    \overline\zeta_\mu \equiv \sum_{\nu\ne\mu} \zeta_\nu \bmod 2.
\end{equation}
Here we use the convenient notation
\be
\gamma_n \,\equiv\, \prod_{\mu=0}^3(\gamma_\mu)^{n_\mu},
\ee
where $n$ is a four-component vector  with $n_\mu\in\mathbb{Z}_2$, to 
label the sixteen independent gamma matrices.

This symmetry means that any low-momentum mode~$\psi(x)$ is exactly 
equivalent to fifteen other modes, $\Bz(x)\,\psi(x)$ for~$\zeta\ne0$. 
These modes are obviously different from the original mode and from 
each other since $\Bz(x)\,\psi(x)$ has very large 
momentum~$p\approx\zeta\pi/a$ if $\psi(x)$ is a low-momentum mode (see 
\eq{Bz-def}). Consequently the single naive-quark field actually 
creates sixteen different flavors or ``tastes'' of quark. The field 
manages this trick by packing each of the different quark tastes into a 
different corner of its Brillouin zone, with taste~$\zeta$ 
corresponding to momenta~$p$ near~$\zeta\pi/a$.

Sixteen is fifteen too many quarks, but the extra degrees of freedom 
are easily removed if the different tastes do not mix (see 
\cite{hisq-paper} for a review). The central issue, therefore, is how 
much mixing there is between  different tastes. The canonical analyses 
address this issue by re-expressing the action in terms of 
interpolating fields for each taste. In the canonical coordinate-space 
analysis, the lattice is divided into hypercubes that have two sites 
per side, and the interpolating fields are defined by weighted averages 
of the naive-quark field over the hypercube. The field corresponding to 
taste~$\zeta$ is~\cite{hisq-paper-appendixB}
\be \label{Psi-def}
\Psi^{(\zeta)}(x_B) \equiv \cS\,
\Bz(x_B) \,\psi(x_B)
\ee
where $x_B$, with $x_{B\mu}/a \bmod 2 = 0$, identifies the hypercube, 
and $\cS$~is a smearing operator defined by
\be
\cS \equiv \prod_\mu \cS_\mu
\ee
with
\be
	\cS_\mu \,\psi(x) \equiv \frac{\psi(x)+\psi(x+a\hat\mu)}{2}.
\ee

The smearing operator is critical to this definition. Its role is most 
easily understood in momentum space, where the operator becomes a 
simple function of momentum~\cite{center-footnote}:
\begin{align}
\cS_\mu(p) &= \e^{ip_\mu a/2}\,\cos(p_\mu a/2) \\
	&\to
	\begin{cases}
		1 & \text{for $p_\mu\to0$} \\
		i\tilde{p}_\mu a/2 & \text{for $p_\mu\to\pi/a + \tilde{p}$}.
	\end{cases} \label{S-approx}
\end{align}
It is used in the definition of~$\Psi^{(\zeta)}$ to suppress 
$p_\mu\approx\pi/a$ while leaving $p_\mu\approx0$ unchanged. To see why 
this is important, we Fourier transform \eq{Psi-def} to obtain
\begin{equation} \label{psi-fft}
\Psi^{(\zeta)}(p) = \sum_{\xi_\mu\in\mathbb{Z}_2} 
\cS(p+\xi\pi/a)\,\gamma_{\,\overline{\zeta}}\,
\psi(p+(\zeta+\xi)\pi/a)
\end{equation}
where we have used the Fourier transform of the doubling 
transformation~\eq{dsym},
\be
	\psi(p) \to \gamma_{\,\overline{\zeta}}\,\psi(p+\zeta\pi/a).
\ee
Momentum~$p$ in~$\Psi^{(\zeta)}(p)$ is restricted to $-\pi/2a < p_\mu 
\le \pi/2a $ since the field is defined on the blocked lattice, which 
has lattice spacing~$2a$; this is also why we need the sum over~$\xi$. 
The terms in~$\Psi^{(\zeta)}(p)$ with~$\xi\ne0$ correspond to tastes 
other than~$\zeta$, but they are suppressed by the smearing function 
which is~$\order(ap)$ unless~$\xi=0$. Consequently,
\be \label{Psi-approx}
	\Psi^{(\zeta)}(p) \to
	\gamma_{\,\overline{\zeta}} \,\psi(p+\zeta\pi/a)
	 \,+\,\order(pa\psi) \quad \text{for $|p_\mu|\ll\pi/a$.}
\ee
The smearing operator, working together with~$\Bz$, serves to isolate 
momenta in the vicinity of $\zeta\pi/a$\,---\,that is, it isolates 
taste~$\zeta$, as is needed for the interpolating field.

It is obvious from this analysis that the isolation of individual 
tastes is only approximate; $\Psi^{(\zeta)}$ is contaminated by other 
tastes,~$\xi+\zeta$, because the smearing operator suppresses these 
tastes by factors of only~$\order(ap)$. Given that $\Psi^{(\zeta)}$, 
\emph{by its definition}, contains other tastes, it seems likely that 
its field equation will exhibit $\order(a)$~taste mixing terms.

We can verify that this is the case by examining a solution $\psi(p)$ 
of the naive-quark Dirac equation,
\be \label{dirac-eq}
	\left(\sum_\mu \frac{i\sin(p_\mu a)}{a} \gamma_\mu + m \right) \psi(p) 
= 0.
\ee
Note that this equation can be rewritten
\be
	\left(\sum_\mu (-1)^{\zeta_\mu}\,\frac{i\sin(p_\mu a)}{a} \gamma_\mu + 
m \right) \psi(p+\zeta\pi/a) = 0.
\ee
or, equivalently,
\be
	\left(\sum_\mu \frac{i\sin(p_\mu a)}{a} \gamma_\mu + m \right) \,
	\gamma_{\,\overline\zeta}\,\psi(p+\zeta\pi/a) = 0,
\ee
which is a restatement of the doubling symmetry.

Consider the blocked field with $\zeta=0$, for example. It is easy to 
show that the Dirac equation,~\eq{dirac-eq}, for $\psi(p)$ implies a 
blocked-lattice Dirac equation for $\Psi^{(0)}(p)$ with a small 
residue:
\be \label{Psi-eq}
\left(\sum_\mu \frac{i\sin(2p_\mu a)}{2a} \gamma_\mu + m \right) 
\Psi^{(0)}(p) = \Delta
\ee
where the residue~$\Delta$ is (obviously)
\begin{align}
	\Delta \equiv \sum_{\xi,\mu} &
	 \left( \frac{i\sin(2p_\mu a)}{2a}
	-  (-1)^{\xi\mu}\,\frac{i\sin(p_\mu a)}{a} \right) \times \notag  \\
	& \quad\times \gamma_\mu \cS(p+\xi\pi/a)
	\psi(p+\xi\pi/a).
\end{align}
Terms with $\xi\ne0$ cause taste mixing, and when $\xi_\mu=1$ these are 
suppressed only to the extent that $\cS(p+\xi\pi/a)$ is small.
In the continuum limit, $|p_\mu|\ll\pi/a$, the largest contributions to 
$\Delta$ therefore arise when only one component of $\xi$ is nonzero, 
$\xi = \hat\mu$, and therefore, using Eqs.~(\ref{S-approx}) 
and~(\ref{Psi-approx}), we find that
\begin{align}
	\Delta &\approx \sum_\mu 
(2ip_\mu)\cS(p+\hat\mu\pi/a)\psi(p+\hat\mu\pi/a)\\
	&\approx - \sum_{\mu=0}^3 \,(-1)^\mu\, ap_\mu^2\, \gamma_5 
\Psi^{(\hat\mu)}(p).
	\label{Delta-approx}
\end{align}
As expected, $\Delta$ contains $\order(a)$~taste mixing.

This $\order(a)$~taste mixing is the conventional textbook 
result~\cite{map-to-stagg}. Our analysis shows that the suppression 
factor of~$(ap)$ comes directly from the smearing operator. It is easy 
to design a new smearing operator for which the suppression is much 
stronger. For example, we can replace $\cS_\mu$ by $\tilde\cS_\mu$
\begin{align}
	\tilde\cS_\mu & \psi(x) \equiv \notag \\ & \frac{\psi(x-a\hat\mu) + 
3\psi(x)+3\psi(x+a\hat\mu) +\psi(x+2a\hat\mu)}{8}
\end{align}
where now
\begin{align}
	\tilde\cS_\mu(p) &= \e^{ip_\mu a/2}\,
	 	\frac{3\cos(p_\mu a/2)+\cos(3p_\mu a/2)}{4} \\
		&\to
		\begin{cases}
			1 & \text{for $p_\mu\to0$} \\
			-i(\tilde{p}_\mu a)^3/8 & \text{for $p_\mu\to\pi/a + \tilde{p}$}.
		\end{cases} \label{Stilde-approx}
\end{align}
With this new smearing, the taste-mixing part of the residue~$\Delta$ 
in the equation for the blocked field (\eq{Psi-eq}) is suppressed by 
order~$(pa)^3$ rather than~$(ap)$:
\be
	\Delta\to\tilde\Delta
	=  \sum_{\mu=0}^3 \,(-1)^\mu\, (a^3p_\mu^4/4)\,
	\gamma_5 \Psi^{(\hat\mu)}(p) + \cdots,
\ee
where we have shown only taste-mixing terms.

Obviously one can suppress taste mixing to even higher orders in the 
lattice spacing by using increasingly nonlocal smearings. Indeed one 
can suppress taste mixing completely, in the noninteracting theory, by 
simply redefining
\be
	\Psi^{(\zeta)}(p) \equiv \gamma_{\,\overline\zeta\,} 
\psi(p+\zeta\pi/a)
	\prod_\mu \theta(|p_\mu|\le\pi/2a).
\ee
In coordinate space this would involve smearing over the entire 
lattice.

This last definition is, of course, the standard momentum-space flavor 
basis. Using this basis it is trivially obvious that taste mixing is 
impossible in the free theory because it would violate momentum 
conservation. Momentum conservation is required by the translation 
symmetry of the underlying theory. Taste-mixing terms appear when we 
use the coordinate-space basis because that basis is defined using a 
hypercubic blocking of the lattice that breaks translation invariance 
and therefore violates momentum conservation, as is evident from the 
$\xi$~sum in~\eq{psi-fft}. The momentum-space flavor basis, on the 
other hand, allows us to analyze separate tastes without violating 
momentum conservation and so is much more useful for studying 
finite-$a$ errors in the theory.

The impossibility of taste mixing in the free theory follows 
immediately from the formal Symanzik analysis of finite-$a$ corrections 
to the full interacting naive-quark action. Any finite-$a$ correction 
to the original action must be a local operator that preserves all of 
the symmetries of the original theory, including gauge symmetry, 
translation invariance, doubling symmetry, parity, and so on. The 
combination of the doubling and translation symmetries implies that 
quark bilinears, in both the interacting and free theories, must have 
flavor-spin structure $\gamma_n\otimes 1$~\cite{gammax1-ref}. For naive 
quarks these bilinears have the form
\be
	\psib\gamma_n\otimes 1\psi \to \psib(x) \gamma_n \psi(x\pm\delta x_n),
\ee
where
\be
	\delta x_n^\mu \equiv  n^\mu\,a
\ee
guarantees the doubling symmetry. Consequently, these operators must be 
taste singlets in the noninteracting theory and cannot mix tastes; 
taste mixing is not allowed in the free theory!

Such quark bilinears can and do mix different tastes when they are 
coupled to other fields, like the gluon field. This happens, for 
example, when the gluon field in a quark-gluon vertex carries off 
momentum~$q\approx\zeta\pi/a$ from the quark line. Such a momentum 
transfer leaves an on-shell quark still on shell, but with a different 
taste. The factor $(-1)^{\zeta\cdot x/a}$ in the gluon field alters the 
taste content of the quark bilinear: in general, an operator
\be
	\psib(x)\gamma_n\psi(x\pm\delta 
x_n)\,\times\,(-1)^{\overline{\zeta}\cdot x/a},
\ee
where the phase comes from a gluon (or other) field coupled to the 
quark bilinear, behaves in effect as though it had spin-taste structure 
$\gamma_{n-\zeta}\otimes\xi_{\zeta}$~\cite{gammax1-ref}. Consequently 
even a simple operator, like $\psib(x)\gamma_\mu 
U_\mu(x)\psi(x+a\hat\mu)$, which has nominal spin-taste 
structure~$\gamma_\mu\times1$, actually has a very complicated 
spin-taste structure since the gluon field can carry off momenta with 
one or more components of order~$\pi/a$.

The gluon in a taste-mixing interaction of this sort is highly virtual 
and so must be reabsorbed almost immediately by the same or another 
quark. Consequently the effects of such a taste-exchange interaction 
are indistinguishable, to first approximation, from a taste-exchange 
four-quark operator. Since four-quark operators are at least dimension 
six, this means taste exchange can arise first in~$\order(a^2)$. Again 
this result follows immediately from the symmetry restrictions on quark 
bilinears in a naive/staggered-quark theory: the only way a quark 
bilinear in a naive-quark action can mix tastes is by coupling to a 
highly virtual gluon or gluons. This is the only mechanism for taste 
mixing allowed by the symmetries of naive/staggered-quark theories.

The doubling symmetry obviously complicates the interpretation of 
correction terms in the quark action when interactions are included. 
Indeed it is more complicated still since the possibility that an 
on-shell quark can emit a highly virtual gluon and yet remain on shell 
means that additional factors of $a\partial_\mu$ acting on the gluon 
field in a quark bilinear do not necessarily suppress the operator by 
additional factors of~$a$ as would normally be the 
case~\cite{lepage2-paper}. Despite these complications, the analysis of 
low-order Symanzik corrections to the naive-quark action is 
straightforward. In~$\order(a)$, the only operators that might enter 
are the standard
\be
a\psib 1\otimes1 D^2\psi \quad\mathrm{and}\quad a\psib 
gF\cdot\sigma\otimes1\psi,
\ee
but each of these is suppressed by an additional factor of~$am$, where 
$m$~is the quark mass, because of the chiral symmetry of the 
naive-quark action when~$m=0$ (see~\cite{hisq-paper}, for example). 
Consequently there are no~$\order(a)$ corrections of any sort; 
$\order(a)$~errors are impossible for gauge-invariant quantities.

The absence of $\order(a)$~errors for physical quantities computed with 
staggered quarks has been known for a long time~\cite{no-a-paper}. This 
should also be true of off-shell quantities that are gauge invariant, 
such as eigenvalues of the naive/staggered-quark Dirac operator, since 
the suppression of $\order(a)$~errors does not rely upon redundancy. 
Redundant operators, which vanish by the equations of motion, have no 
effect on physical quantities but can shift off-shell amplitudes. While 
some of the $\order(a)$~operators are redundant, none of them 
contributes in~$\order(a)$ because they all violate the action's chiral 
symmetry and so are separately suppressed by an additional~$am$. In the 
specific case of the Dirac-operator eigenvalues, furthermore, it is 
hard to see how the eigenvalues could have $\order(a)$~errors without 
causing similar errors in vacuum polarization contributions to physical 
quantities, since the quark determinant in the path integral is the 
product of these eigenvalues. There is no way that such errors could be 
cancelled by valence-quark effects since these depend differently on 
the number of quark flavors~$n_f$. Consequently the absence 
of~$\order(a)$~errors in physical quantities requires that such errors 
are absent in the eigenvalues as well, as we expect. Recent simulation 
results for the eigenvalues also indicate that errors are 
likely~$\order(a^2)$ and smaller~\cite{eig-paper}.

Strictly speaking, the Symanzik analysis is justified only for a 
straight simulation with naive quarks, including all sixteen tastes (or 
staggered quarks, with four tastes). After extensive study, all 
evidence thus far strongly suggests that the $\order(a^2)$ taste-mixing 
in the sixteen-flavor theory translates into $\order(a^2)$~errors in 
the ``rooted'' 
theory~\cite{sharpe-paper,Kronfeld:2010aw,Donald:2011if}, where the 
quark determinant is replaced by its $1/16^\mathrm{th}$~root in order 
to reduce the effective number of tastes to one. In particular no 
mechanism has been identified that could change an $\order(a^2)$~error 
in the unrooted theory into an $\order(a)$~error in the rooted theory. 
There is certainly no way that the $\order(a)$~term from the canonical 
coordinate-space flavor basis (\eq{stagg-a-term}) could reappear since 
it violates momentum conservation and so was impossible to begin with; 
and there are no other unsuppressed $\order(a)$~operators.

The leading taste-mixing effects, therefore, are from the exchange of 
gluons with momenta of order $\zeta\pi/a$, where 
$\zeta_\mu\in\mathbb{Z}_2$. Luckily these, and all higher-order 
taste-mixing operators are highly perturbative, and so can be 
rigorously analyzed and systematically removed. We have already removed 
both the leading and next-to-leading order taste-mixing in going from 
naive/staggered quarks to ASQTAD quarks, and, more recently, from 
ASQTAD to HISQ quarks~\cite{hisq-paper}. The net effect has been to 
reduce taste-mixing effects by almost an order of magnitude at current 
lattice spacings, as is immediately evident from nonperturbative 
calculations of the mass differences between pions of different 
taste~\cite{smearing-makes-a-irrelevant}. Taste mixing is clearly under 
control.

We acknowledge very useful comments from Steve Sharpe, and grant 
support from the National Science Foundation and Cornell University.


\begin{thebibliography}{23}
\bibitem{center-footnote} The overall phase in $\cS_\mu(p)$ does not 
result in finite~$a$ errors. It disappears if the blocked field is 
associated with the center of the hypercube rather than~$x_B$, one of 
its corners.
\bibitem{map-to-stagg} With only a little extra work, one can show 
that \eq{Delta-approx} is exact if $ap_\mu^2$ is replaced by 
$(1-\cos(2pa))/2a$, which is the same momentum dependence found in the 
canonical staggered-quark analysis. It is also easy to show that the 
spin-taste structure here is $\gamma_5\otimes\xi_5\xi_\mu$, just as in 
the staggered-quark analysis; the relation between naive-quark and 
staggered-quark tastes given in~(C7) of~\cite{hisq-paper} is useful for 
showing this.
\bibitem{gammax1-ref} This follows immediately from Eq.~(E2) in 
Appendix~E of~\cite{hisq-paper}. This equation shows the most general 
quark bilinear that is invariant under doubling transformations. Only 
those operators with no factor of~$(-1)^{\overline{s}\cdot x/a}$ (i.e., 
those with $s=\overline{s}=0$) are translation invariant and therefore 
acceptable in a naive-quark action.
\bibitem{smearing-makes-a-irrelevant} See~\cite{hisq-paper} for recent 
data. The very large suppression of taste splitting in the pions 
resulted from smearing of the gluon operators in the quark action. Such 
smearing obviously suppresses the exchange of highly virtual gluons, 
but has no impact on the $\order(a)$~terms like that 
in~\eq{stagg-a-term}. The fact that smearing the gluons dramatically 
reduced the observed taste splittings is striking evidence that gluon 
exchange is the only mechanism for taste mixing.
\end{thebibliography}
\end{document}